\documentclass[aps,pre, twocolumn,groupedaddress,showpacs]{revtex4-1}
\usepackage{amsmath}
\usepackage{mathtools}
\usepackage{wasysym}
\usepackage{color}
\usepackage{tikz}
\usepackage{xcolor}
\usepackage{soul}
\usepackage{amssymb}
\usepackage{stackengine}
\usepackage{scalerel}
\usepackage{graphicx,color}

\usetikzlibrary{shapes.geometric}
\definecolor{r}{rgb}{1,0,0}
\definecolor{g}{rgb}{0,1,0}
\definecolor{b}{rgb}{0,0,1}

\begin{document}
\bibliographystyle{apsrev}

\newcommand{\markerone}{\raisebox{0.5pt}{\tikz{\node[draw,scale=0.6, circle, fill=black!10!green](){};}}}
\newcommand{\markertwo}{\raisebox{0.5pt}{\tikz{\node[draw,scale=0.6, regular polygon, regular polygon sides = 6, fill=black!10!magenta,rotate=90](){};}}}
\newcommand{\markerthree}{\raisebox{0.5pt}{\tikz{\node[draw,scale=0.4, regular polygon, regular polygon sides = 3, fill=black!10!cyan,rotate=180](){};}}}
\newcommand{\markerfour}{\raisebox{0.5pt}{\tikz{\node[draw,scale=0.6, regular polygon, regular polygon sides = 4, fill=black!10!blue,rotate=0](){};}}}
\newcommand{\markerfive}{\raisebox{0.5pt}{\tikz{\node[draw,scale=0.4, regular polygon, regular polygon sides = 3, fill=black!10!blue,rotate=45](){};}}}
\newcommand{\markersix}{\raisebox{0.5pt}{\tikz{\node[draw,scale=0.4, regular polygon, regular polygon sides = 3, fill=black!10!yellow,rotate=0](){};}}}
\newcommand{\markerseven}{\raisebox{0.5pt}{\tikz{\node[draw,scale=0.4, regular polygon, regular polygon sides = 4, fill=black!10!red,rotate=0](){};}}}
\newcommand{\markereight}{\raisebox{0.5pt}{\tikz{\node[draw,scale=0.4, diamond, fill=black!10!black,rotate=0](){};}}}

\newcommand{\openbigstar}[1][0.7]{
	\scalerel*{
	\stackinset{c}{-.125pt}{c}{}{\scalebox{#1}{\color{blue}{$\bigstar$}}}{$\bigstar$}
	}{\bigstar}
	}

\title{Experimental investigation of water distribution in two-phase zone during gravity-dominated evaporation}



\author{Cesare M. Cejas$^{1,\dagger}$, Jean-Christophe Castaing$^{2}$,  Larry Hough$^{1}$, Christian Fr$\acute{e}$tigny$^{3}$, and R$\acute{e}$mi Dreyfus$^{1}$}
\affiliation{$^{1}$Complex Assemblies of Soft Matter, CNRS-Solvay-UPenn UMI 3254, Bristol, PA 19007-3624, USA}
\affiliation{$^{2}$Solvay Research and Innovation Centre - Aubervilliers, France 93300}
\affiliation{$^{3}$ Sciences et Ing$\acute{e}$nierie de la Mati$\grave{e}$re Molle CNRS SIMM UMR 7615 ESPCI, Paris, France 75005}
\affiliation{$^{\dagger}$ Microfluidics, MEMS, Nanostructures Laboratory, CNRS Gulliver UMR 7083, Institut Pierre Gilles de Gennes (IPGG), ESPCI Paris, PSL Research University, 6 rue Jean Calvin 75005 Paris, France}

\date{\today}

\begin{abstract}
We characterize the water repartition within the partially saturated (two-phase) zone (PSZ) during evaporation out of mixed wettable porous media by controlling the wettability of glass beads, their sizes, and as well the surrounding relative humidity. Here, Capillary numbers are low and under these conditions, the percolating front is stabilized by gravity. Using experimental and numerical analyses, we find that the PSZ saturation decreases with the Bond number, where packing of smaller particles have higher saturation values than packing made of larger particles. Results also reveal that the extent (height) of the PSZ, as well as water saturation in the PSZ, both increase with wettability. We also numerically calculate the saturation exclusively contained in connected liquid films and results show that values are less than the expected PSZ saturation. These results strongly reflect that the two-phase zone is not solely made up of connected capillary networks, but also made of disconnected water clusters or pockets. Moreover, we also find that global saturation (PSZ + full wet zone) decreases with wettability, confirming that greater quantity of water is lost via evaporation with increasing hydrophilicity.  These results show that connected liquid films are favored in more hydrophilic systems while disconnected water pockets are favored in less hydrophilic systems. 
\end{abstract}


\maketitle


Evaporation from porous media such as soil involves complex mechanisms that enable water flow across liquid-air interfaces in the pores. These liquid networks control the rigidity of porous materials~\cite{Herminghaus05}, as well as the total amount of water that plants can extract from the soil~\cite{Calvet88, Cejas14b}. Evaporation affects energy balance and involves phase changes and fluid displacement~\cite{Shaw87} at pore scales thereby providing different levels of challenges when predicting evaporation using theoretical models~\cite{Shokri09}. The challenges are due to the different mechanisms involved at the pore scale level, such as pore impregnation and drying~\cite{Prat07}, liquid flows due to pore size distributions~\cite{Shokri09, Lehmann08} and flows due to local gradients of temperature~\cite{Huinink02}. Because of this inherent complexity, several models and experimental approaches have been established to analyze evaporation, such as pore-network models~\cite{Laurindo98, Yiotis10} and silica spheres sandwiched between glass slides~\cite{Shaw87}. 

Experimental studies on evaporation have been commonly performed in two-dimensions (2D) for simplicity and reproducibility~\cite{Prat99, Yiotis12a}. The morphology of the front and the dynamics of the liquid films during evaporation are affected by a competition between viscous, capillary, and gravity forces~\cite{Shokri08, Prat99, Yiotis12a}. Some studies have probed the effect of reduced gravity conditions on front dynamics and the fluid displacement patterns by using parabolic flight experiments~\cite{Or09} or tilting the sample cell at an angle relative to the direction of gravity~\cite{Yiotis12a}.  Notable studies have also focused on the drying front~\cite{Shokri08} and the fractal formation that accompanies it ~\cite{Yiotis06, Prat99, Prat12}. Most evaporation experiments have been devoted to measuring and predicting the extent or depth of the capillary film network in the two-phase zone~\cite{Shokri08, Yiotis12a}, which links the fully wet region and the evaporating surface~\cite{Prat99}, arguing that comprehension of the capillary film network would be useful for numerous applications. For instance, recent results used evaporation principles on the prediction of total mass loss from combined evapotranspiration fluxes for root growth applications~\cite{Cejas14b}. Some drying studies have also coupled experiments with numerical simulations~\cite{Han13}.

Prediction of the extent of the liquid film region has already been largely studied using various models. However, what still remains an area of interest is the water repartition within the partially saturated (two-phase) zone (PSZ), also referred to as unsaturated zone~\cite{Shokri08} or vadose zone~\cite{Hunt08} in literature. The PSZ is a dynamic air/water mixture formed by saturated pores, liquid bridges, and films that coalesce to form networks, which establish hydraulic connections inside the porous medium~\cite{Yiotis12b}. The negative curvature of the water-air menisci in these films~\cite{Xu08} mean that they are subject to intense pressure that play a crucial role in fluid transport mechanisms, including having an impact on infiltration~\cite{Cejas14a} and groundwater recharge~\cite{Hunt08}. Experimental evidence has revealed that some liquid films are indeed disconnected from the principal network and thus form isolated water clusters or pockets~\cite{Shokri08, Prat99}. Although these isolated clusters have been qualitatively observed in experiments~\cite{Shokri08, Shokri09, Prat99}, our understanding of the evaporation process would benefit from the analysis of the water content repartition within the two-phase zone, i.e. the quantity of water contained in continuous liquid films and those contained in disconnected isolated pockets. Such information will not just contribute to our comprehension of evaporation dynamics but it will help improve applications in soil additive design~\cite{Wei14, Wei13}.

In this paper, we present an experimental investigation of evaporation from a 2D porous medium that consists of a monolayer stack of monodisperse beads inside a Hele-Shaw cell. We study evaporation using well-controlled experimental conditions including different relative humidities, sizes of glass beads, and mixtures of particles with varying wettability. The primary objective is to estimate the general repartition and saturation of water within the two-phase zone (PSZ) subject to gravity. We find evidence of a complex coexistence of connected capillary networks and disconnected water clusters or pockets, where the quantity of the latter dominates during later evaporation times and in less hydrophilic media. We find that the saturation of the PSZ is a function of the Bond number, where packing of smaller particles (lower Bond number) have higher saturation values. From these scalings, we extract information on total saturation in the two-phase zone. By comparing total PSZ saturation values to the saturation exclusively contained in continuous films, we obtain the amount of water contained in isolated pockets. We find that for different wetting conditions, the increase in hydrophobicity in the porous medium is generally accompanied by an increase in the amount of water in isolated pockets.

\section{Materials and Methods}

\textbf{Cell preparation}. We perform experiments in Hele-Shaw cells constructed by putting together two glass sheets (Peerless Store and Glass, PA) of $15$cm in length and $10$cm in width. The thickness of the cell is the diameter of the beads, $e = d = 2R$. Because of the silicon needed to glue the sheets together, the effective size of the medium becomes $13.5$cm in length and $7.5$cm width. The glass sheets are coated with hydrophobic silane solution (OMS Chemicals, Canada) to reduce wetting effects.  When oriented vertically, only the top of the cell is left unsealed. Once dry, the cells are filled with glass beads forming a monolayer. We study two different glass bead diameters:  $d_1 = 0.5 \pm 0.1$mm (soda lime, MO-SCI) and $d_2 = 1.0 \pm 0.1$mm (borosilicate, Sigma-Aldrich). For each experiment,  we measure the cross-section of the cell, $S$. The thickness of the Hele-Shaw cell is $e_1 = 0.53 \pm 0.05$mm and $e_2 = 1.1 \pm 0.1$mm. As a result, our system is quasi-2D.

\begin{table}
\caption{Properties of water at $T$=25$^{\circ}$C}
\resizebox{8.0cm}{!}{
\begin{tabular}{c c c c}
	\hline
	{\textbf{Quantity}} & {\textbf{Symbol}} & {\textbf{Unit}} & {\textbf{Value}} \\
	\hline
	Diffusion of water in air & $D_{M}$ & m$^2$/s & $2.5$$\cdot$$10^{-5}$ \\
	Diffusion of water in porous medium & $D_{eff}$ & m$^2$/s & $D_M \phi^{1.5}$ \\
	Liquid density & $\rho$ & kg/m$^3$ & $1000$ \\
	Surface tension & $\sigma$ & mN/m & $73$ \\
	\hline
\end{tabular}
}
\end{table}

We use deionized water, whose properties are shown in Table(1). The porosity of the medium, $\varphi$, is experimentally measured for the entire cell using standard imbibition method. The average porosities for $d_1$ and $d_2$ are $\varphi_1=0.63 \pm 0.02$ and $\varphi_2=0.53 \pm 0.02$ respectively. 

\textbf{Surface treatment}. We use a heterogeneous mixture, $\alpha$, of glass beads of two different wettabilities, characterized by contact angles: (1) more hydrophilic, $\theta_{1} = 16 \pm 5^{\circ}$, for glass beads washed with  0.01M HCl for few hours at $70^{\circ}$C and neutralized by washing with deionized water and left to dry in the oven at $70^{\circ}$C for $24$ hours; and (2) less hydrophilic, $\theta_{2} = 82 \pm 4^{\circ}$ for glass beads washed with the same hydrophobic silane solution. We use the Cassie equation (Eq.(1)) to estimate the equivalent contact angle, $\theta_{equ}$, of our mixture. 
\begin{equation}
   cos(\theta_{equ}) = \alpha(cos\theta_1) + (1 - \alpha)(cos\theta_2),
\label{Eq1}
\end{equation}
Eq.(1) is based on the principle that a mixture of particles having two different contact angles will have a macroscopic effective contact angle that is proportional to their respective fractional components, $\alpha$~\cite{Shokri08}. This is a reasonable assumption in the framework of the experimental conditions since the beads are essentially hydrophilic, only one type is less hydrophilic than the other. The values are presented in Table(2).

\begin{table}
\caption{Equivalent contact angle of porous media mixture, $\alpha$, of $\theta_{1} = 16 \pm 5^{\circ}$ and $\theta_{2} = 82 \pm 4^{\circ}$. }
\resizebox{5.0cm}{!}{
\begin{tabular}{c c c c c c}
	\hline
	$\alpha$ & $0$ & $0.25$ & $0.50$ & $0.75$ & $1$  \\
	\hline
	$\theta_{equ}$ ($^{\circ}$) & $82$ & $68.6$ & $56.2$ & $41.6$ & $16$ \\
	\hline
\end{tabular}
}
\end{table}

\begin{figure}
\includegraphics[width=3in]{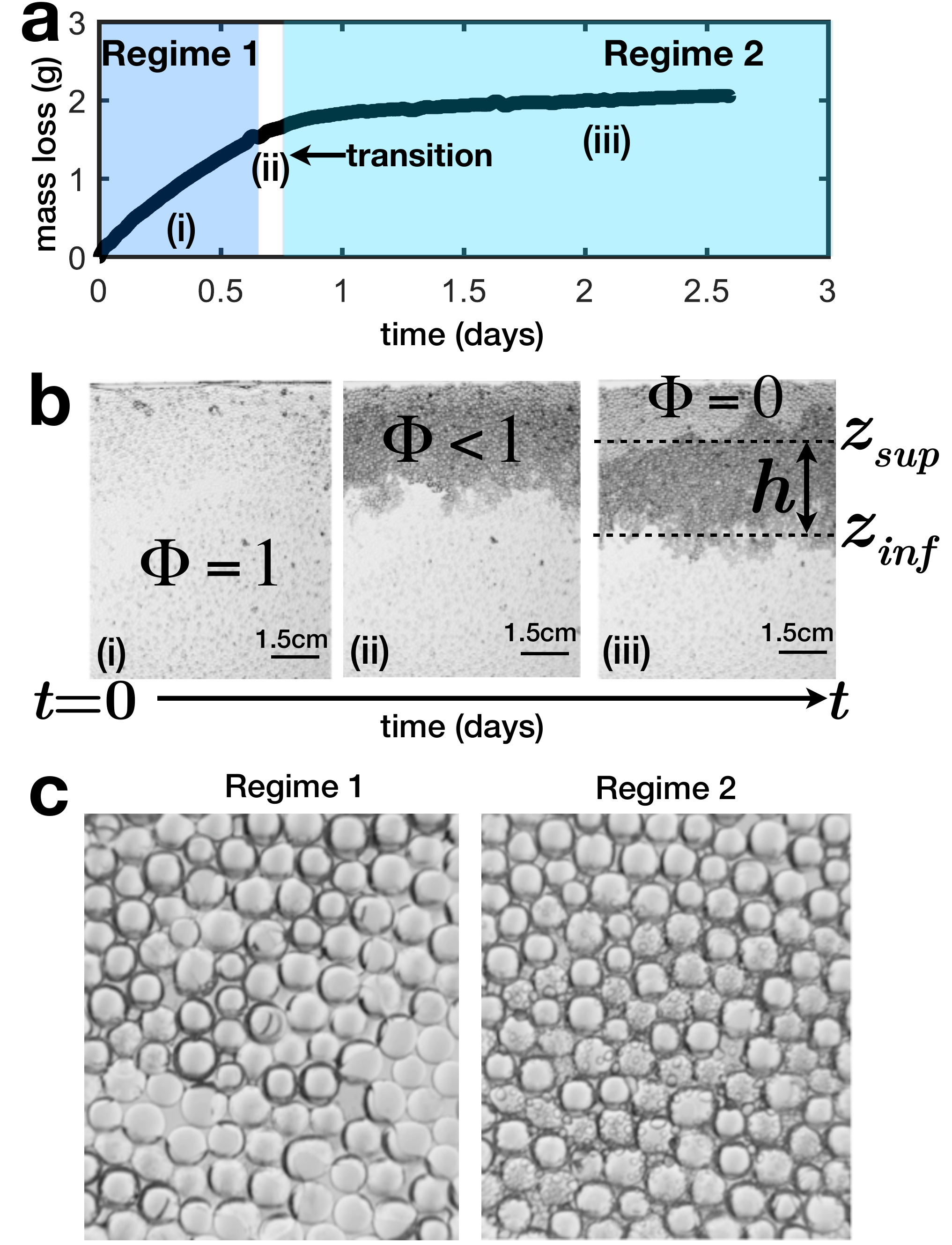}
\caption{({\textbf{a}}) Typical evaporation curve showing mass loss, $m=\widetilde{m}(t)$, described by two regimes separated by a transition point. ({\textbf{b}}) Experimental images of the time evolution of the water distribution with three distinct zones: dry zone ($\Phi = 0$), partially saturated zone (PSZ) ($\Phi < 1$), and a fully saturated or wet zone ($\Phi = 1$). The PSZ is bound by an evaporating front, $z_{sup}$ (in the upper portion/superior portion), and a percolating front, $z_{inf}$ (in the lower/inferior portion). Scale bar refers to $\approx$1.5cm. ({\textbf{c}}) Experimental image during Regime 1 showing liquid films in the form of pendular bridges between adjacent beads. The second image is typical of Regime 2 showing the appearance of disconnected droplets, which can coalesce to form water pockets.}
 \label{Fig1}
\end{figure}

\textbf{Experimental conditions}. We fill the cells with water by fully submerging them in a closed water tank attached to a vacuum pump, which draws air out of the medium as the surrounding water slowly percolates through the pores. The pump also reduces the presence of bubbles that could otherwise affect evaporation measurements. Once filled with water, the cell is oriented vertically such that water evaporates from the top. Evaporation experiments are performed inside a glove box with humidification systems, which permit precise control of the humidity. The relative humidity, $H_R$, is regulated and set at $20.0 \pm 2.0\%$ and $40.0 \pm 2.0\%$ by servo-control device (model 5100, Electro-Tech Systems) and recorded. The temperature inside the glove box remains constant at $T = 32 \pm 2^{\circ}$C. We measure the mass loss of water using a digital balance (P4002 model, Denver Instruments) at specific time intervals controlled by a Matlab program. 

\textbf{Image treatment}.  We obtain images of the water repartition by taking pictures every $30$min using a Canon 500D camera with $18-55$-mm lens with a spatial resolution of about $35$~$\mu$m per pixel. The cells are back-illuminated allowing for the observations of the three zones, as shown in Fig.(1)a : the dry zone, the partially saturated zone (PSZ), and the fully saturated zone. Images are treated using Matlab

\section{Definitions and Experimental Data}

\textbf{Observations and Flux Definition}. We perform evaporation experiments using a wide variety of physical parameters (particle size, wettability fraction, humidity) and each experiment exhibits similar features. A typical mass loss curve, $m=f(t)$, due to evaporation is shown in Fig.(1)a. The corresponding images are shown in Fig.(1)b. Let $\Phi$ be the saturation level of the medium. The system starts out fully saturated, $\Phi=1$ (Fig.(1)b). Evaporation proceeds via two regimes. During the first regime, air enters the medium and displaces water resulting to the appearance of a two-phase air-liquid mixture called the partially saturated zone (PSZ), where $0<\Phi<1$. The PSZ consists of an intricate hydraulic network connecting $z_{inf}$ and $z_{sup}$. The first regime is controlled by mass transfer across an external mass transfer length scale, $\delta$, above the cell surface, where drying rates~\cite{Coussot00, Shokri08, Lehmann08, Yiotis12a} are approximately constant and mass loss due to evaporation is linear with time~\cite{Chauvet09, vanBrakel80}. During regime 1, rate of mass loss remains high due to hydraulic connections that ensure rapid transport of water from the bulk (interior of the porous medium) to the surface. As a result, water is continuously supplied at the surface to sustain the rate of evaporation~\cite{Coussot00}, an action that is macroscopically described as a wicking effect~\cite{Yiotis12a}. In literature, regime 1 is also referred to as the constant rate period (CRP)~\cite{Yiotis12a, Prat99, Lehmann09, Chauvet10, Shokri08}.

At a certain characteristic depth of $z_{inf}$, the liquid films can no longer sustain capillary flow and they detach from the evaporating surface. The disruption of the hydraulically connected films marks the transition, (Fig.(1)a), from the first to the second regime characterized by much lower evaporation rates~\cite{Schultz91}. At this point, a dry region, $\Phi=0$ now separates the front and the evaporating surface (Fig.(1)b). Due to loss of hydraulic connections at the surface, evaporation during the regime 2 now shifts to vapor diffusion~\cite{Shokri09} across air-filled pores in the dry region and drying rates continue to slow down. In literature, regime 2 is also referred to as the falling rate period (FRP)~\cite{Yiotis12a, Prat99, Lehmann09, Chauvet10, Shokri08}.

The presence of the two regimes signifies that the overall evaporation flux can be separated into $J_1$ (flux regime 1) and $J_2$ (flux regime 2). Analytical expressions of both fluxes can be found extensively in literature~\cite{Yiotis12a, Yiotis12b, Prat99, Chauvet09, Chauvet10, Shokri08, Cejas14b}, showing that the mass loss in regime 1 is linear with time while the mass loss in regime 2 scales as $\sqrt{t}$. Such descriptions are in accordance with our own experimental observations.

\begin{figure}
\includegraphics[width=3.5in]{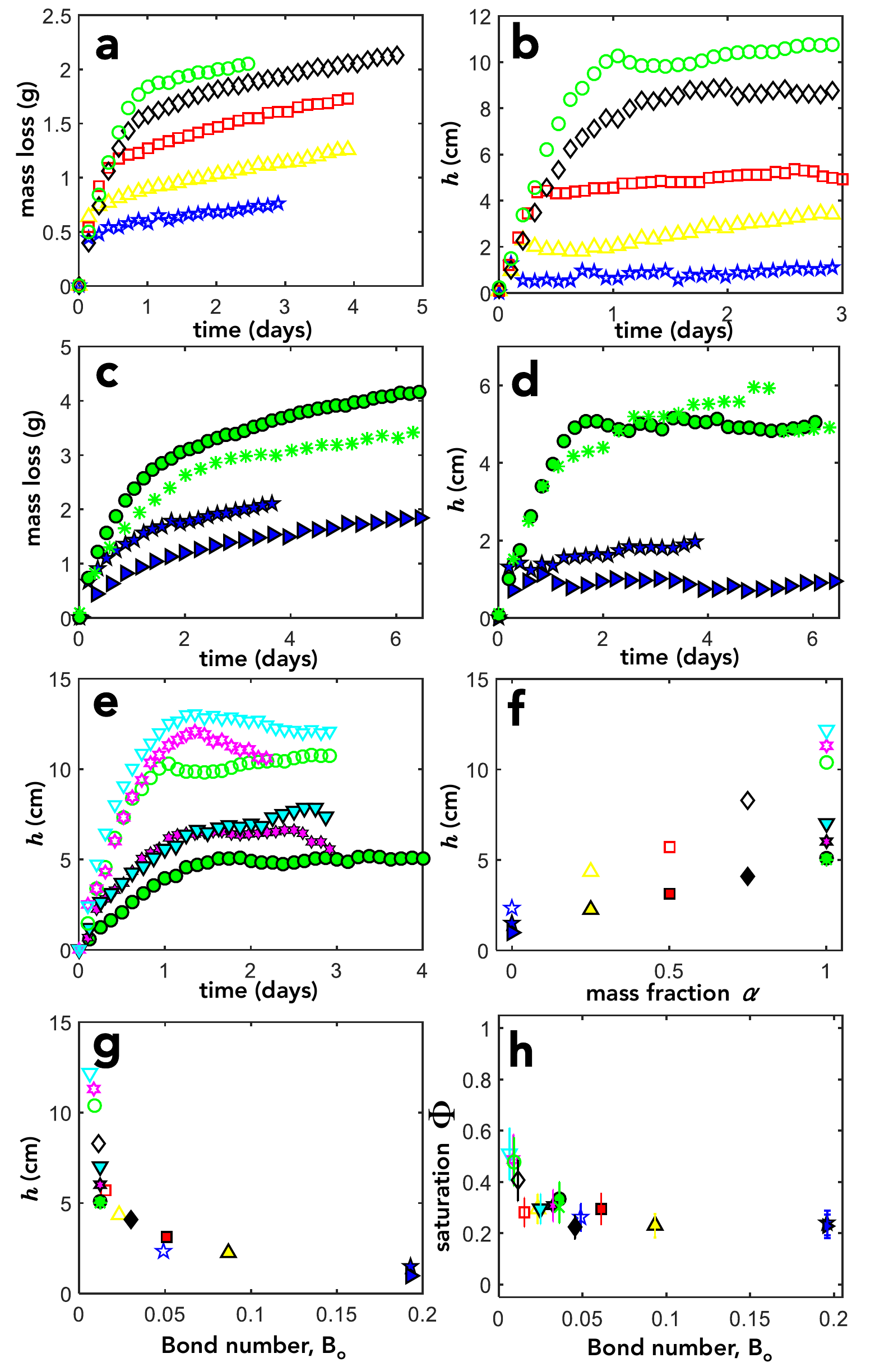}
\caption{({\textbf{a}}) $m=\widetilde{m}(t)$ for $d_1 = 500\mu$m; ({\textbf{b}})  $h=\widetilde{h}(t)$ for $d_1 = 500\mu$m;  ({\textbf{c}}) Experimental curves for $d_2 = 1$mm of $m=\widetilde{m}(t, H_R)$ for $H_R = 20.0 \pm 2.0\%$ and $H_R = 40.0 \pm 2.0\%$.  ({\textbf{d}}) Experimental curves of $h=\widetilde{h}(t, H_R)$ for two different $H_R$ at $d_2=1$mm. ({\textbf{e}}) $h=\widetilde{h}(t)$ for $d_1$ and $d_2$ for different $\theta_t$. ({\textbf{f}}) $h=\widetilde{h}(\alpha)$ for $d_1$ and $d_2$ where $h$ increases with $\alpha$. ({\textbf{g}}) $h=\widetilde{h}(B_o)$ for $d_1$ and $d_2$. ({\textbf{h}}) Experimentally measured PSZ saturation, $\Phi_{Expt}=\widetilde{\Phi}(B_o)$, for $d_1$ and $d_2$.  ({\textbf{Legend}}): All performed at $H_R = 20\%$, $\theta_t=90^{\circ}$, $d_1$ (hollow points): ({\color{blue}{$\star$}}) $\alpha=0$; ({\color{yellow}{$\bigtriangleup$}}) $\alpha=0.25$; ({\color{red}{$\Box$}}) $\alpha=0.50$; ({\color{black}{$\diamondsuit$}}) $\alpha=0.75$; ({\color{green}{$\circ$}}) $\alpha=1$.  All performed at $H_R = 20\%$, $\theta_t=90^{\circ}$, $d_2$ (filled points): ($\openbigstar$) $\alpha=0$; ($\protect\markersix$) $\alpha=0.25$; ($\protect\markerseven$) $\alpha=0.50$; ($\protect\markereight$) $\alpha=0.75$; ($\protect\markerone$) $\alpha=1$.  All performed at $\alpha=1$ and $H_R = 20\%$: ({\color{magenta}{$\davidsstar$}}) $d_1$, $\theta_t=75^{\circ}$; ({\color{cyan}{$\bigtriangledown$}}) $d_1$, $\theta_t=45^{\circ}$; ($\protect\markertwo$) $d_2$, $\theta_t=75^{\circ}$; ($\protect\markerthree$) $d_2$, $\theta_t=45^{\circ}$. All performed at $H_R = 40\%$ and $d_2$: ($\protect\markerfive$) $\alpha=0$; ({\color{green}{$\varhexstar$}}) $\alpha=1$. }
 \label{Fig2}
\end{figure}

In our experiments, we control the wettability by varying the ratio, $\alpha$. We show in Fig.(2)a, $m = \widetilde{m}(t, \alpha)$, for particle size $d_1$. Results for $d_2$ are shown in SI-1. From these results, greater mass of water is lost in more hydrophilic systems than in less hydrophilic systems since more hydrophilic particles maintain hydraulic continuity that favors evaporation. This is consistent with reported experimental~\cite{Shokri09} and simulation results~\cite{Chraibi09}. The minimal $h$ value observed in less hydrophilic systems is attributed to the minimal presence of hydraulic continuity between the bulk and the surface~\cite{Shokri09}. 

The height or extent, $h$, of the PSZ is bound by the positions of the percolating front, $z_{inf}$, and the evaporating front, $z_{sup}$.The $h$ value is a function of wettability mixture, $h = \widetilde{h}(t, \alpha)$, whose results are presented in Fig.(2)b ($d_1$) and SI-1 ($d_2$). From the figures, the different regimes are easy to distinguish, where $h$ increases linearly during regime 1 but remains approximately constant during regime 2. 

We also control the relative humidity, $H_R$, of the glove box. The curves in Fig.(2)c and Fig.(2)d show $m=\widetilde{m}(t, H_R)$ and $h=\widetilde{h}(t, H_R)$ respectively for $H_R = 20.0 \pm 2.0\%$ and $H_R = 40.0 \pm 2.0\%$ for $m=\widetilde{m}(t, H_R)$ using $d_2= 1$mm. The relative humidity is a measure of the amount of moisture in the air. At low $H_R$, air is considered dry and thus water evaporates faster to maintain equilibrium. At higher $H_R$ values, there is greater moisture content in the air and thus evaporation rates proceed slower. The extent of $h$ is similar for the two different relative humidity values. This strongly suggests that $h$ is independent of relative humidity. The relative humidity only changes the overall water content (Fig.(2)c) inside the medium but not the extent, $h$, of the liquid film region (Fig.(2)d). 

In an evaporation experiment, water transport is driven and retarded by various forces. We therefore estimate the dimensionless Bond, $B_o$, and Capillary numbers, $C_a$, to determine which forces are dominant. 


\textbf{Dimensionless numbers}. The capillary forces that transport water towards the evaporating surface are always opposed by viscous forces~\cite{Laurindo98, Prat02, Lehmann08, Lehmann09, Yiotis12a}. Because in these experiments the sample cells are oriented vertically, gravity also has an opposing against upward capillary flow.  When gravity is part of drying, the pressure of the liquid is hydrostatic and the displacement pattern follows features of invasion percolation~\cite{Prat99, Yiotis10, Prat12}.

The interplay between these forces controls the dynamic aspects governing evaporation. To determine the relative dominance among the forces, we compute $C_a$ and $B_o$, which provide the ratio between viscous and capillary forces (former) and between gravity and capillary forces (latter) and are defined as:
\begin{equation}
   C_a = \frac{\mu v}{\sigma \text{cos}{\theta_{equ}}}~\text{and}~{B_o} = \frac{\rho g R^2}{\sigma \text{cos}{\theta_{equ}}},
\label{Eq5}
\end{equation}
where $\mu$ is the dynamic liquid viscosity, $v$ is the characteristic liquid filtration velocity, $\sigma$ is the interfacial tension, $g$ is the acceleration due to gravity. The parameter $v$ in $C_a$ is normally defined as the filtration velocity in regime 1 and is represented as, $v = J_1/\rho$~\cite{Prat02}. While maintaining the same $R$, we can tune the value of $B_o$ by tilting the sample cell at an angle relative to the direction of gravity, henceforth referred to as a tilt angle, $\theta_t$. This effectively reduces gravity into a component given by $g \sin{\theta_t}$, where its effect on the extent of the PSZ, $h$, is shown in Fig.(2)e. Lower $\theta_t$ values result to higher $h$ values. The experimental values for $J_1$, $B_o$, and $C_a$ are presented in Table(3).

\begin{table}
\caption{Values of $J_1$, Bo, and $Ca$ for different experimental conditions. }
\resizebox{7.5cm}{!}{
\begin{tabular}{c c c c c c c c c c}
	\hline
	Exp & $R$ & $H_R$ & $\theta_t$ & $\alpha$ & $J_1$ & Bo & $C_a$  \\
	No. & ($10^{-6}$m) & ($\%$) & $^{\circ}$ & no unit & g/day & ($10^{-2}$) & ($10^{-8}$) \\
	\hline
	1 & 250 & 20 & 90 & 0 & 2.75 & 4.90 & 4.93 \\
	2 & 250 & 20 & 90 & 0.25 & 2.56 & 2.33 & 3.37 \\
	3 & 250 & 20 & 90 & 0.50 & 2.40 & 1.53 & 1.85 \\
	4 & 250 & 20 & 90 & 0.75 & 1.85 & 1.14 & 1.06 \\
	5 & 250 & 20 & 90 & 1 & 1.86 & 0.91 & 0.85  \\
	& & & & & & & & & \\
	6 & 500 & 20 & 90 & 0 & 1.35 & 19.6 & 3.33  \\
	7 & 500 & 20 & 90 & 0.25 & 1.63 & 9.33 & 1.91  \\
	8 & 500 & 20 & 90 & 0.50 & 1.54 & 6.12 & 1.19  \\
	9 & 500 & 20 & 90 & 0.75 & 1.99 & 4.55 & 1.14  \\
	10 & 500 & 20 & 90 & 1& 1.96 & 3.62 & 0.89  \\
	& & & & & & & & & \\
	11 & 500 & 40 & 90 & 0 & 0.72 & 19.6 & 3.33  \\
	12 & 500 & 40 & 90 & 1 & 1.14 & 3.62 & 0.89 \\
	& & & & & & & & & \\
	13 & 250 & 20 & 75 & 1 & 2.48 & 0.87 & 1.13 \\
	14 & 250 & 20 & 45 & 1 & 3.45 & 0.64	& 1.57  \\
	15 & 500 & 20 & 75 & 1 & 3.00 & 3.50 & 1.37  \\
	16 & 500 & 20 & 45 & 1&  3.86 & 2.56 & 1.71  \\
	\hline 
\end{tabular}
}
\end{table}

A summary of $h$ results for all experiments is shown in Fig.(2)f as a function of $\alpha$, revealing that smaller particles, $d_1$ (hollow points), produce bigger $h$ values than larger particles, $d_2$ (filled points). In addition, Fig.(2)f also shows that $h$ increases with hydrophilic fraction $\alpha$. In Fig.(2)g, all the data collapse on a single curve as $h$ is plotted as a function of $B_o$, revealing a sharp decrease. When plotted in log-log scale, the regression analysis shows a slope of $-0.6$ (see SI-2), thereby giving a scaling law of $h \sim B_o~^{-0.6}$. Furthermore, we calculate $\Phi_{Expt}$, which is the ratio of the mass of water within the PSZ with respect to the initial mass of water, $m_i$, at the start of the experiment, while taking into account the geometry of the cell:
\begin{equation}
   	\Phi_{Expt} = \Phi_{PSZ} = \frac{m_f - (S \varphi \rho L^{*})}{m_i},
\label{Eq3}
\end{equation}
where $m_f$ is the equilibrium total mass of water in the porous medium, which can be estimated from the mass curves, $L^{*} = L - z_{inf}$, where $L$ is the vertical length of the medium from the cell surface to the bottom of the packing, and $\varphi$ is porosity. The plot of PSZ saturation $\Phi_{Expt}$ as function of $B_o$ is shown in Fig.(2)h. Four principal observations can be deduced from these data: (1) smaller particles hold more water content than larger particles; (2) PSZ saturation increases slightly with wettable fraction; (3) PSZ saturation is higher when gravity effects are diminished; and (4) PSZ saturation appears to be constant regardless of relative humidity.


\textbf{Continuous films, isolated pockets}. Experimental images of the PSZ in Fig.(1)c show varying compositions at different regimes. Fig.(1)c (left) shows a network of liquid films, which are a coalescence of pendular bridges between beads, trapped in pore spaces. Studies~\cite{Yiotis12a, Prat07} have shown that liquid films pinned to the corners of the pore wall (corner film flow) contain menisci with varying radii of curvature that induce a capillary pressure gradient. This serves as the driving force for upward liquid flow from the bulk to the evaporating surface, thus assuring hydraulic continuity. In Fig.(1)c (right), we observe droplets with positive curvature, which suggests they have been detached from the main film network. Most of the droplets start appearing later (during regime 2). Apart from droplets, disconnected films also appear because as water content in the liquid films continues to decrease due to evaporation, they eventually detach from the main liquid network.

To understand water repartition in the PSZ, we must first quantify the saturation of water contained in continuous liquid films, which contribute to extent of $h$.

\textbf{Corner film flow}.Yiotis $et~al$~\cite{Yiotis12a} developed a model that predicts $h$ and the saturation of liquid films in the PSZ during evaporation out of a porous medium by calculating the cross-section of the corner liquid films. These corner films, when viewed from the top, resemble flow in a non-circular capillary tube~\cite{Dong95, Chauvet09} shown in Fig.(3)a. Studies~\cite{Camassel05,Chauvet09, Chauvet10, Prat02} have demonstrated that evaporation of liquid from a square tube results to the formation of corner liquid films (Fig.(3)b), whose large cross-sectional area reduces flow resistance from the interior of the porous medium (bulk) to the evaporating surface. The corner films represent a more effective transport mechanism than simple vapor diffusion~\cite{Camassel05}, which is the dominant mechanism in the evaporation of liquid from a circular tube. Transport via corner film flow holds true for any tube of polygonal cross-section with sharp corners~\cite{Prat07, Yiotis12a}. 

As liquid films begin to thin out towards the end of regime 1, the geometry of the films then adopt a more rounded shape during regime 2~\cite{Yiotis12b} (Fig.(3)b), thereby decreasing cross-sectional area and increasing flow resistance from the bulk to the evaporating surface. The flow resistance along the corner film is also a function of contact angle~\cite{Ransohoff88}, where higher contact angles reduce the cross-sectional area of the film available for flow.

{\setlength\intextsep{0pt}
\begin{figure}[tp] 
\includegraphics[width=3.5in]{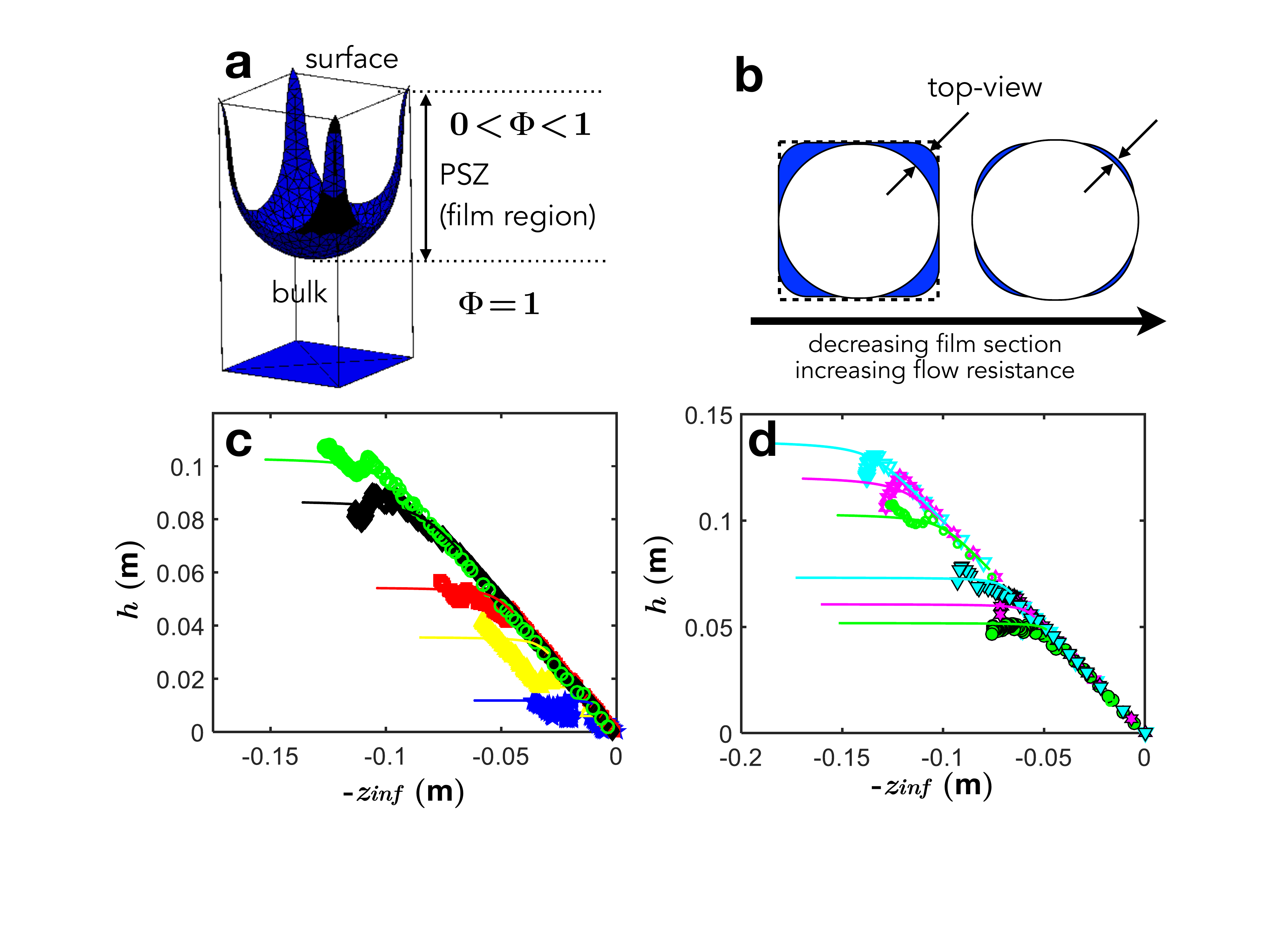}
\caption{({\textbf{a}}) Scheme of corner liquid films of a square tube during evaporation. Image is from Surface Evolver simulations using the code of Berthier and Brakke~\cite{Berthier12}. ({\textbf{b}}) Top-view of liquid film thickness in a tube of square cross-section. Corner films in pore walls are initially sharp but takes a more rounded shape~\cite{Yiotis12b} as evaporation proceeds. ({\textbf{c}}) Experimental results (points) of $h=\widetilde{h}(z_{inf})$ for $d_1$ for different $\alpha$ fitted with the model (solid lines) of Yiotis $et$~$al$~\cite{Yiotis12b}. Comparison for $d_2$ is shown in SI-1. ({\textbf{d}}) Experimental results (points) of $h=\widetilde{h}(z_{inf})$ for $d_1$ and $d_2$ but performed at varying tilt angles, $\theta_t$. Solid lines show the model of Yiotis $et$~$al$~\cite{Yiotis12b}. The extent of $h$ is higher at smaller $\theta_t$. Legend definitions are the same ones used in Fig.(2). }
 \label{Fig3}
\end{figure}

\section{Results and Discussion}

\textbf{Prediction of PSZ height}. To quantify the water saturation contained in liquid films, we first must calculate for $h$ using the analytical model of Yiotis $et~al$~\cite{Yiotis12a, Yiotis12b} with data from Dong and Chatzis~\cite{Dong95}. The model is straightforward, widely covered in literature, and detailed discussion on its use is not the primary objective of this paper. We then compare their model with our experiments shown in Fig.(3)c for $d_1$ ($d_2$ in SI-1) and Fig.(3)d for different $\theta_t$. Comparison reveals excellent agreement. Apart from predicting $h$, their model also permits calculation of the saturation of water contained in liquid films in the PSZ but had been only developed for completely hydrophilic materials. Due to our mixed wettable system, we pursue a different avenue, through numerical simulations, to find out the quantity of water contained in connected liquid films and from that information, we deduce the quantity of water disconnected from the main network (droplets, pockets, etc). 

\textbf{Simulations}. We define the total water saturation within the PSZ in Eq.(3). We also define the saturation contained in continuous liquid films and saturation isolated from the main network, both with respect to initial mass, by performing numerical simulations using Surface Evolver~\cite{Brakke10}. We consider $h$ to be primarily determined by a series of interconnected pendular bridges between particles in contact with each other or with the wall. In a 2D granular system of mixed wettability, we can have different kinds of pendular bridges in Fig.(4)a: bridge between beads of the same $R$ and of the same surface wetting properties, (i) HI-HI and (ii) HO-HO, where HI stands for more hydrophilic particles while HO stands for less hydrophilic particles; (iii) bridge between beads of the same $R$ but of different wetting properties, HI-HO; (iv) bridge between a more hydrophilic bead and the less hydrophilic wall, HI-HOwall; (v) and finally a bridge between a less hydrophilic bead and the less hydrophilic wall, HO-HOwall.  The shape of the meniscus depends on the wetting properties of the solids (sphere/plane) and also on the volume of the liquid body.

It is understandably difficult to assess the liquid bridges during regime 1, due to the rapid nature of its flux. Thus, in these simulations, we consider that the bridge is in equilibrium during the slower evaporation dynamics in regime 2 - a fact supported by results in Fig.(2) showing the quasi-constant value of $h$. Earlier, we estimate the $B_o$ $\sim 10^{-2}$ (see Table(3)). At equilibrium, the capillary force is the dominant force acting on the bridge since the liquid volumes are very small.  

The menisci of pendular bridges have negative curvatures, whose pressure is hydrostatic along the extent of $h$. The pressure value, $P$, is determined using Surface Evolver~\cite{Berthier12, Brakke10} by calculating change in energy with respect to volume, $P=dE/dV$. The results from the numerical simulations establish a pressure-volume relationship, which can be used to evaluate the saturation of water contained in liquid bridges within the PSZ. Since a liquid film is, in essence, a coalescence of capillary bridges, it is then important to estimate the number of bridges that comprises this network in the PSZ as a function of wettability. 

In the case of random-close packing, where the distribution is close to dense packing, each spherical bead has a maximum of six neighbors. Thus, the minimum number of contact sites between two beads is $3 N_b$ while there are $2 N_b$ contact sites between a bead and the wall as shown in Fig.(4)f, where $N_b$ is the number of spheres. For a certain $S$ and $h$, the parameter $N_b$ can be estimated using principles associated with Kepler conjecture, which describes the maximum packing of spheres (or objects) achieved for a given space~\cite{Hales05}. We then evaluate the saturation contained in continuous liquid films, $V_C$, along the extent of $h$, using the equation:
\begin{equation}
\begin{split}
   	V_C &= \frac{3 N_b}{h} (\alpha^2 \Sigma_{II} + \alpha(1-\alpha)\Sigma_{IO} + (1 - \alpha)^2\Sigma_{OO}) \\
		& +  \frac{2 N_b}{h} (\alpha \Sigma_{I-wall} + (1 - \alpha)\Sigma_{O-wall}),
\label{Eq4}
\end{split}
\end{equation}
where the various $\Sigma$ parameters refer to:
\begin{equation}
\begin{split}
   	 \Sigma_{II} &= \int^h_0\, V_{II}(z) dz, \\
   	 \Sigma_{IO} &= \int^h_0\, V_{IO}(z) dz, \\
	\Sigma_{OO} &= \int^h_0\, V_{OO}(z) dz, \\
	\Sigma_{I-wall} &= \int^h_0\, V_{I-wall}(z) dz, \\
	\Sigma_{O-wall}&= \int^h_0\, V_{O-wall}(z) dz, \\
\label{Eq5}
\end{split}
\end{equation}
where $V_{II}$, $V_{IO}$, $V_{OO}$, $V_{I-wall}$, and $V_{O-wall}$ correspond to the volumes of the liquid trapped between two solid bodies. The subscripts refer to the surface coupling between two different wetting properties: ``I'' for more hydrophilic and ``O'' for less hydrophilic. The $h$ values are those predicted by the model of Yiotis $et$~$al$~\cite{Yiotis12a, Yiotis12b}, which we have shown to agree with experiments.

{\setlength\intextsep{0pt}
\begin{figure}
\includegraphics[width=3.5in]{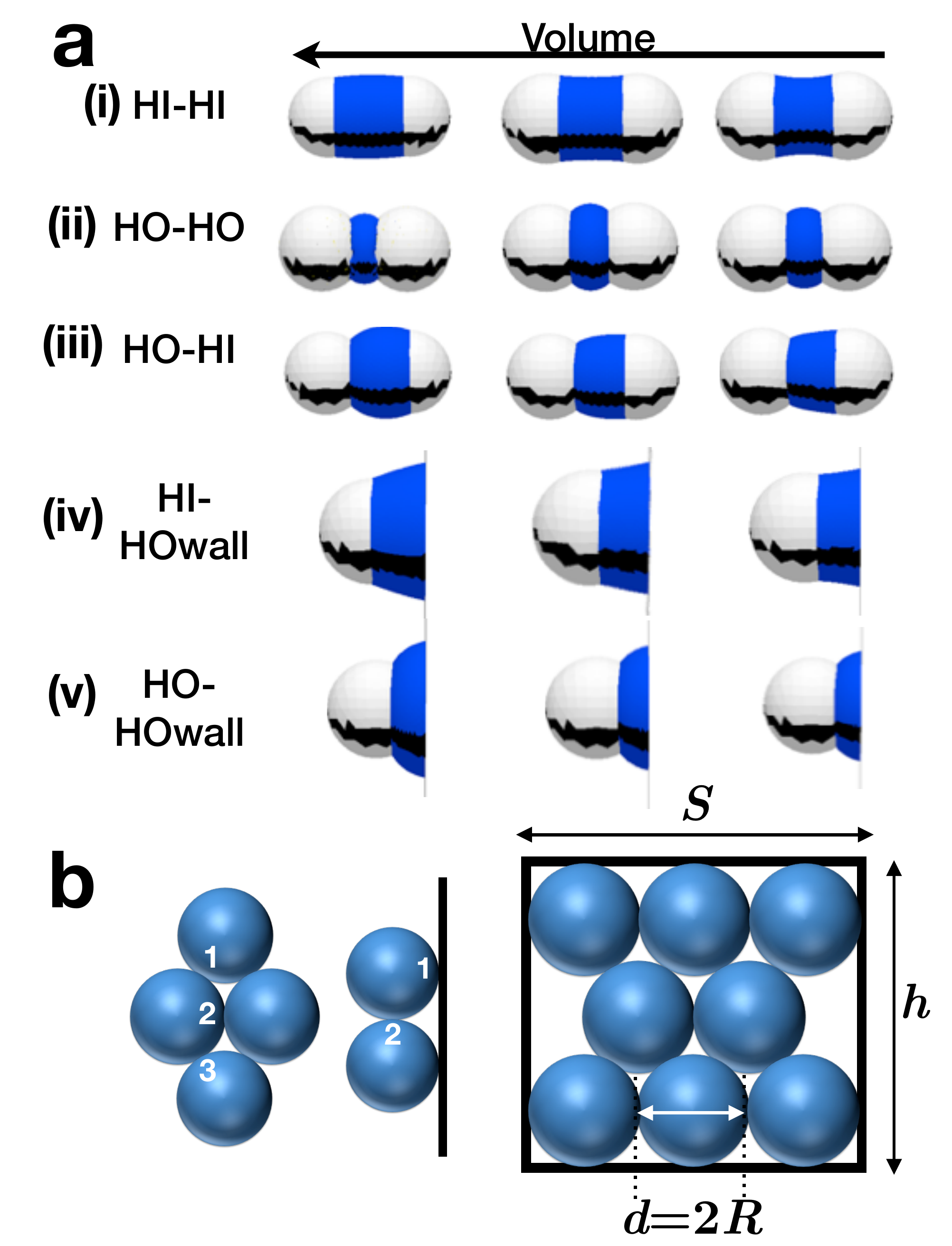}
\caption{({\textbf{a}}) Surface Evolver results showing the evolution of bridge meniscus between two spherical bodies and between a sphere and a vertical plane of varying surface properties. ({\textbf{i}}) HI-HI; ({\textbf{ii}}) HO-HO; ({\textbf{iii}}) HI-HO; ({\textbf{iv}}) HI-HOwall; and ({\textbf{v}}) HO-HOwall, where HI refers to more hydrophilic surface while HO refers to less hydrophilic surface. ({\textbf{b}}) Illustration of the number of contact sites of a given sphere adjacent to other spheres or to a wall. Representation of the number of spheres that can be contained within a given space. }
 \label{Fig4}
\end{figure}

The total saturation of the liquid films in the PSZ can be obtained from simulation expressed in terms of $\Phi_{Evolver}$ defined as:
\begin{equation}
   	\Phi_{Evolver} = \frac{V_C}{V_{empty}},
\label{Eq6}
\end{equation}
where $V_C$ is the volume of the continuous liquid films calculated from Eq.(6), $V_{empty}$ is the free pore volume of the entire medium defined as $V_{empty} = S L \varphi$. Eq.(6) shows the saturation of connected liquid films within the PSZ, taking note that the fully wet-zone has no liquid-air interfaces. We show in Fig.(5)a the plot of $\Phi_{Expt}$ (Eq. (3)) as function of $B_o$. The black line is a guide for the eye.
 
{\setlength\intextsep{0pt}
\begin{figure}
\includegraphics[width=3.5in]{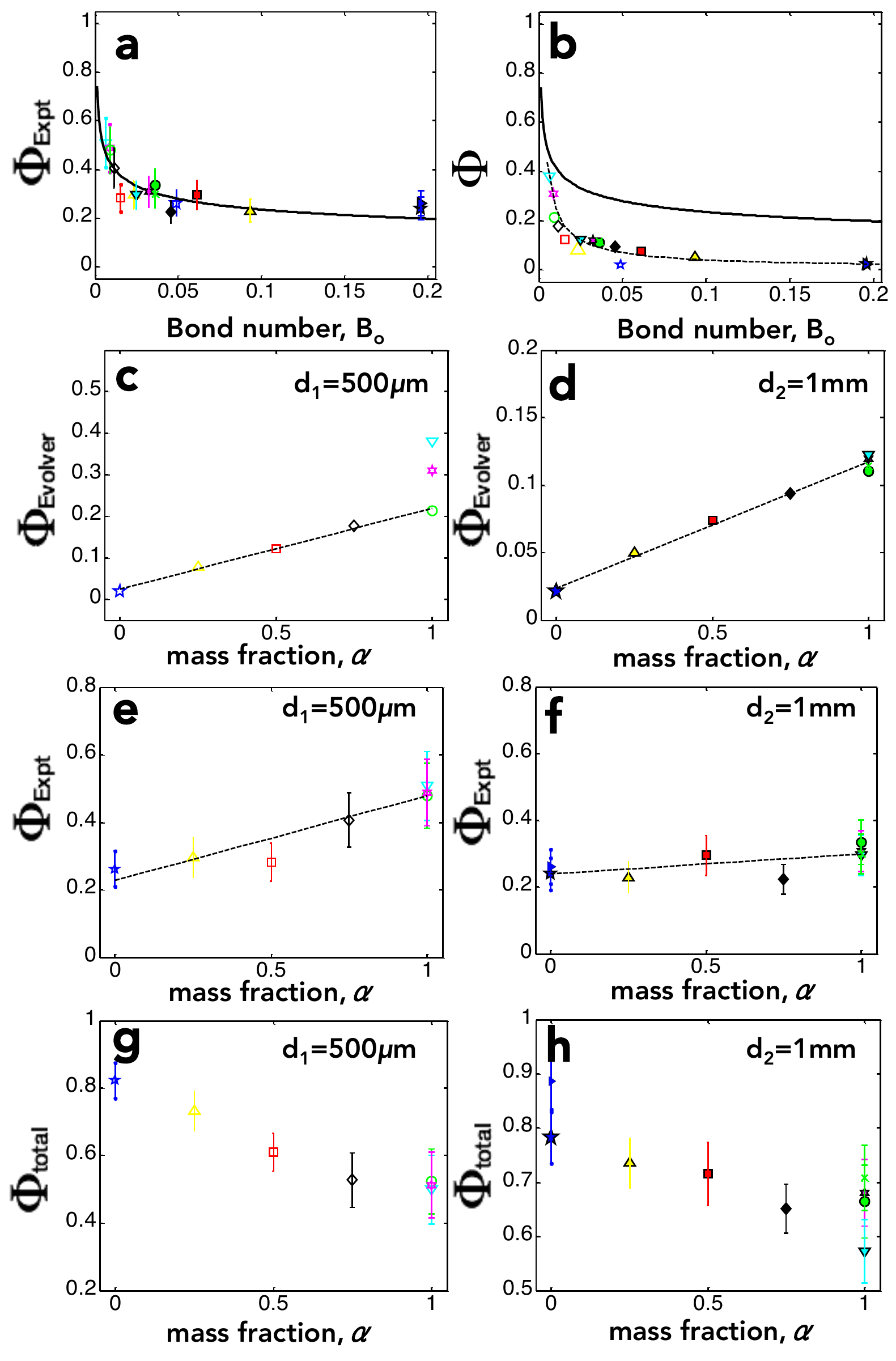}
\caption{({\textbf{a}}) Saturation of the PSZ, $\Phi_{Expt}$, as a function of Bond number, $B_o$, where solid line is a guide for the eye. ({\textbf{b}}) Plot of simulation results from Surface Evolver, $\Phi_{Evolver}$ (points), as function of $B_o$ compared to $\Phi_{Expt}$ (solid line). The broken line is the best fit of the results (SI-3) of the analytical model~\cite{Yiotis12a, Yiotis12b} and serves as a guide. ({\textbf{c}}) Plot of $\Phi_{Evolver}$ as function of wettability, $\alpha$, for $d_1$. Broken line serves as guide. ({\textbf{d}}) Plot of $\Phi_{Evolver}$ as function of $\alpha$ for $d_2$. Broken line serves as guide. ({\textbf{e}}) Plot of $\Phi_{Expt}$ as function of $\alpha$ for $d_1$. Broken line serves as guide.({\textbf{f}}) Plot of $\Phi_{Expt}$ as function of $\alpha$ for $d_2$. Broken line also serves as guide. ({\textbf{g}}) Plot of $\Phi_{total}$ (total saturation including $\Phi_{Expt}$ and saturation within fully wet zone) as function of $\alpha$ for $d_1$. ({\textbf{h}}) Plot of $\Phi_{total}$ as function of  $\alpha$ for $d_2$.  Legend definitions in all graphs are the same ones used in Fig.(2).}
 \label{Fig5}
\end{figure}

We also calculate the saturation contained in liquid films using the model of Yiotis, $et$~$al$~\cite{Yiotis12a, Yiotis12b} (graph in SI-3), which was conceptualized for completely hydrophilic media and from a single tube. Thus, the value obtained from this model was then scaled with respect to 2D porous medium, which can be seen as a bundle of equivalent capillary tubes~\cite{Lehmann08}. We then compare in the main figure in Fig.(5)b the results of the analytical model (now shown as a broken line) with results from the simulations (points), $\Phi_{Evolver}$ as function of $B_o$. The broken line in this figure is the best fit from the results of the analytical model~\cite{Yiotis12a, Yiotis12b} and serves as a guide (see SI-3). From these results, the fact that there is an offset in saturation contained in liquid films with respect to the experimentally measured PSZ saturation (solid line) strongly reflects the fact that the PSZ is not solely made up of connected hydraulic networks but also contains water in the form of droplets and disconnected pockets, which have actually also been observed using tomography~\cite{Shokri08, Shokri09}. 

Results in Fig.(5)c and Fig.(5)d first show that $\Phi_{Evolver}$ increases with $\alpha$, suggesting that more hydrophilic materials favor liquid films. Secondly, comparison of Fig.(5)e and Fig.5(f) show that packing of smaller particles hold more saturation than packing of larger ones due to the higher capillary pressures emanating from smaller liquid volumes. Both figures also show the PSZ saturation seems to slightly increase with $\alpha$, notably for smaller diameters, similar to the variation of $h$ with $\alpha$ in Fig. 2(f). This shows that increasing hydrophilicity favors saturation  in continuous films while increasing hydrophobicity favors disconnected pockets. Thirdly, PSZ saturation is generally higher at greater $\theta_t$ because the extent $h$ also increases with reduced gravity effects. Lastly, PSZ saturation seems to be independent of relative humidity. This may appear misleading; however, if we take into account the fully wet zone (FSZ), which is larger at higher $H_R$, the overall water content remaining in the entire medium would still be greater at higher $H_R$, in accordance with observations in Fig.(2)a. In Fig.(5)g and Fig.5(h), we show $\Phi_{total}$ (PSZ + FSZ) as function of wettability. Greater quantity of water is globally retained in less hydrophilic media in contrast to more hydrophilic ones.

\section{Conclusion}

We have systematically investigated experiments on the evaporation of water from model soil systems using parameters such as surface treatment, particle size, and environmental humidity. We demonstrate through experiments and simulations that the partially saturated zone is not only a connected liquid film region but also a region containing droplets and clusters that have been disconnected from the main hydraulic network. We calculate the average global saturation in the partially saturated zone and find that the saturation of water in the continuous liquid films increases with overall hydrophilicity since they favor formation of capillary networks. In contrast, less hydrophilic systems do not favor liquid film continuity and thus greater quantity of disconnected clusters exist. The more disconnected clusters present in the porous medium, the less we lose water in the medium since evaporation now mainly proceeds via diffusion instead of capillary transport through the network.

These 2D results could have direct applications in soil additive design to control water repartition in soil during evaporation. Soil is essentially a mixture of hydrophilic and hydrophobic components.  One way to achieve an optimal retention of water is to develop additives with dual wettable properties. The presence of hydrophobic (or less hydrophilic) components breaks liquid film continuity and will result to more disconnected clusters, which evaporate at slower rates. These experiments have been performed in 2D model systems. It would be a good contribution to the subject in the future to analyze the connectivity of films and appearance of liquid clusters in three dimensional systems, since such systems most closely resemble real soil structures.



\begin{acknowledgments}
We thank the support of Solvay, CNRS, and UPenn. We thank everyone in COMPASS (UMI 3254) laboratory, as well as Zhiyun Chen, Jennifer Reiser, and Prof. Douglas J. Durian (UPenn) for helpful exchanges. We are also very grateful to K. Brakke for helping us with Surface Evolver.
\end{acknowledgments}




\end{document}